\documentclass[conference]{IEEEtran}
\IEEEoverridecommandlockouts
\usepackage{cite}
\usepackage{amsmath,amssymb,amsfonts}
\usepackage{algorithmic}
\usepackage{graphicx}
\usepackage{textcomp}
\usepackage{xcolor}
\allowdisplaybreaks[3]

\def\BibTeX{{\rm B\kern-.05em{\sc i\kern-.025em b}\kern-.08em
    T\kern-.1667em\lower.7ex\hbox{E}\kern-.125emX}}

\begin{document}

\title{Secure OFDM-IM ISAC With Artificial-Noise-Aided Index Deception\\
\thanks{This work was carried out in part during the research visit of Ufuk Altun to the Institute of Communications Engineering at TU Dortmund University, hosted by Onur Günlü. The research visit was supported by the Research Explorer Ruhr (RER) program of the Research Academy Ruhr. The work of Onur Günlü was partially supported by the ZENITH Research and Leadership Career Development Fund under Grant ID23.01, EU COST Action 6G-PHYSEC,  Swedish Foundation for Strategic Research (SSF) under Grant ID24-0087, and German Federal Ministry of Research, Technology and Space (BMFTR) 6GEM+ Research Hub under the Grants 16KIS2412 and 16KISS005.}
}

\author{\IEEEauthorblockN{Ufuk Altun}
\IEEEauthorblockA{\textit{Department of Electrical and Electronics Engineering} \\
\textit{Koç University,}
Istanbul, Turkey \\
ualtun20@ku.edu.tr}
\and
\IEEEauthorblockN{Onur Günlü}
\IEEEauthorblockA{\textit{Lehrstuhl für Nachrichtentechnik} \\
\textit{TU Dortmund University,} Dortmund, Germany \\
onur.guenlue@tu-dortmund.de}
}

\maketitle

\begin{abstract}
We propose an artificial noise (AN)-aided, secure OFDM with index modulation (OFDM-IM) framework for integrated sensing and communication (ISAC). Consider a security-critical scenario, where the sensing target is also a potential eavesdropper. Instead of suppressing the signal power toward the target, the proposed method injects AN into inactive OFDM-IM subcarriers. The precoder spatially nulls the AN at the legitimate receiver, while projecting active-subcarrier energy levels toward the target to deceive its index detector. Since the monostatic ISAC transmitter knows the AN waveform, the same inactive subcarriers also contribute to sensing. We formulate this scenario as an optimization problem that maximizes transmit power toward the target while providing secure communication with a legitimate receiver. Numerical results show that the proposed scheme improves targeted transmit power compared to OFDM.
\end{abstract}

\begin{IEEEkeywords}
Integrated sensing and communication (ISAC), OFDM with index modulation (OFDM-IM), physical layer security, artificial noise, index deception, secure sensing, semidefinite relaxation (SDR).
\end{IEEEkeywords}

\section{Introduction}

Integrated sensing and communication (ISAC) is a key enabling technology for sixth-generation (6G) wireless networks, where data transmission and radar sensing are jointly supported by a shared spectral and hardware platform \cite{Wei2022,Qaisar2026}. By reusing communication waveforms for environmental sensing, ISAC can improve spectral efficiency and support emerging applications such as vehicular networks, low-altitude platforms, and intelligent transportation systems. However, this dual-functional operation also introduces new physical-layer security (PLS) challenges \cite{Gunlu2023,Zhu2025,Matsumine2025, Zhou2025}. In particular, a sensing target may simultaneously act as an untrusted eavesdropper. In this case, the transmitter must illuminate the target for sensing while preventing the same receiver from reliably extracting confidential communication information \cite{Su2021,Su2022}.

Several secure ISAC designs have addressed this tension through AN-aided beamforming, interference exploitation, sensing-assisted eavesdropper localization, robust secrecy beamforming, sensing-privacy protection, and RIS-assisted waveform optimization \cite{Su2021,Su2022,Su2024,Cao2025,Zou2024,Meng2025}. These works show that spatial and covariance-domain degrees of freedom can degrade unauthorized reception while maintaining sensing functionality. However, existing secure ISAC designs do not exploit the index-domain structure of index-modulated multicarrier waveforms, where information is also embedded in the subcarrier activation pattern.


Orthogonal frequency division multiplexing with index modulation (OFDM-IM) is a multicarrier transmission technique in which information is conveyed not only by conventional constellation symbols, but also by the indices of active subcarriers \cite{Basar2013}. Due to its sparse activation pattern, OFDM-IM has been studied in the context of jamming, eavesdropping, and adaptive PLS \cite{Kaplan2020,Altun2022a,Altun2024}. In parallel, OFDM-IM and related index-modulated waveforms have also been investigated for ISAC systems, where the active-subcarrier pattern provides an additional waveform degree of freedom for sensing and communication \cite{Sahin2021,Hawkins2024,Singh2023,Temiz2023}. However, index structure of OFDM-IM is vulnerable to eavesdropping since an attacker can attempt to recover the active-subcarrier set through index detection.

We propose an AN-aided secure OFDM-IM ISAC framework that directly targets this index-detection vulnerability. The key idea is to inject structured AN into the data-inactive subcarriers of the OFDM-IM. These subcarriers remain inactive with respect to data transmission and preserve the OFDM-IM information-bearing structure, but they are reused as security and sensing resources. The AN is spatially precoded so that it is nulled at the legitimate receiver. At the same time, it is directed toward the target/eavesdropper and designed to appear as active-subcarrier-like energy. When the eavesdropper is also the sensing target, this approach mitigates the conventional conflict between sensing and communication performance. The main contributions are summarized as follows:
\begin{itemize}
\item We introduce an AN-aided OFDM-IM ISAC architecture in which data-inactive subcarriers are reused as structured AN-bearing sensing and security resources.

\item We formulate a transmit covariance optimization problem that jointly enforces reliable communication at the legitimate receiver, AN suppression at the legitimate receiver, and improved transmit power toward the target.

\item We show that, unlike a classical zero-forcing (ZF)-secured OFDM baseline, the proposed scheme can achieve a higher transmit power toward the target/eavesdropper while degrading Eve's index detection.

\item We evaluate the proposed design using three complementary metrics: target-direction transmit power, spatial beampatterns, and index-detection error probability.

\end{itemize}

\section{System Model}
\label{sec:system_model}
\begin{figure*}
\centering
\includegraphics[width=1\linewidth]{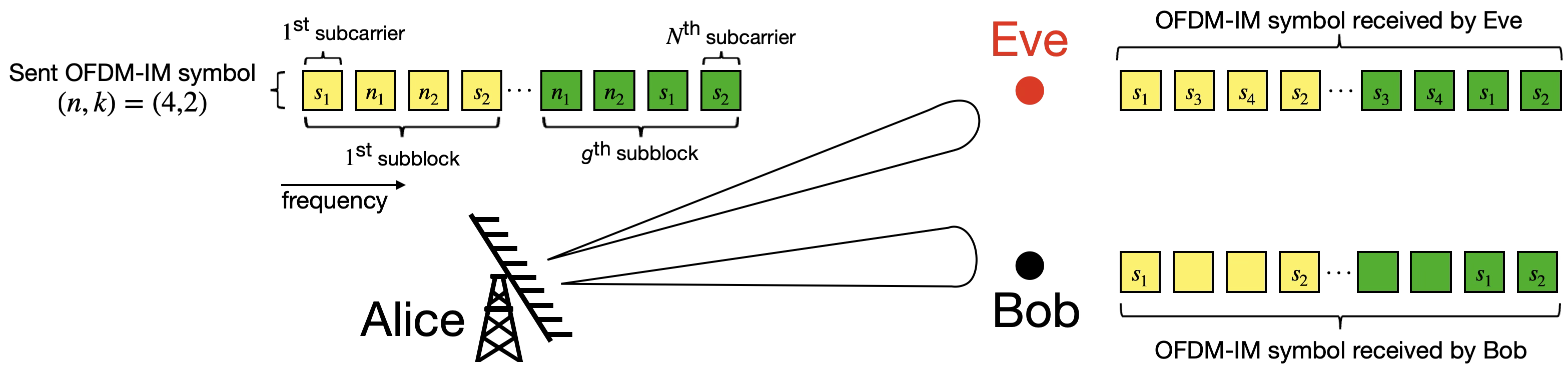}
\caption{Proposed AN-aided OFDM-IM ISAC system model. The BS transmits a dual-functional waveform in which data-inactive subcarriers are reused for spatially precoded AN. Through Bob-side AN suppression, the legitimate receiver observes the intended sparse OFDM-IM structure for index decoding, while the target/eavesdropper observes misleading active-subcarrier-like energy on data-inactive subcarriers.}
\label{fig:system_model}
\end{figure*}

We consider a multi-antenna OFDM-IM-based ISAC system, as illustrated in Fig.~\ref{fig:system_model}. A base station (BS), denoted by Alice, is equipped with $N_t$ transmit antennas and communicates with a single-antenna legitimate receiver, denoted by Bob. At the same time, Alice illuminates a sensing target located at angle $\theta_t$. The target is assumed to be a potential passive eavesdropper, denoted by Eve, which attempts to infer the legitimate OFDM-IM information. This target/eavesdropper model captures a security-critical ISAC scenario in which suppressing the signal toward Eve may directly reduce the sensing performance.

\subsection{OFDM-IM Subcarrier Structure}

Alice employs an OFDM-IM waveform with $N$ subcarriers. The subcarriers are divided into $G$ subblocks, each containing $n$ subcarriers, such that $N=Gn$. The set of subcarrier indices in the $g$-th subblock is denoted by $\mathcal{K}_{g}$.
In each subblock, $k$ out of $n$ subcarriers are selected as data-active subcarriers, while the remaining $(n-k)$ subcarriers are data-inactive. The data-active and data-inactive subcarrier sets in the $g$-th subblock are denoted by $\mathcal{K}_{g,a}$ and $\mathcal{K}_{g,p}$, respectively, where
\begin{equation}
\mathcal{K}_{g} = \mathcal{K}_{g,a} \cup \mathcal{K}_{g,p}, \qquad \mathcal{K}_{g,a} \cap \mathcal{K}_{g,p} = \emptyset,
\end{equation}
with $|\mathcal{K}_{g,a}|=k$ and $|\mathcal{K}_{g,p}|=n-k$.
The active-subcarrier pattern conveys
\begin{equation}
p_1 = \left\lfloor \log_2 \binom{n}{k} \right\rfloor
\end{equation}
index bits, while the $k$ active constellation symbols convey
\begin{equation}
p_2 = k\log_2 M
\end{equation}
symbol bits, where $M$ is the modulation order. Hence, each OFDM-IM subblock conveys $p=p_1+p_2$ bits. Let
\begin{equation}
\mathcal{C}_{g} = \left\{ \mathcal{C}_{g,1}, \ldots, \mathcal{C}_{g,2^{p_1}} \right\}
\end{equation}
denote the OFDM-IM index codebook for the $g$-th subblock, where $\mathcal{C}_{g,m}\subseteq\mathcal{K}_{g}$ is the $m$-th active-subcarrier pattern and $|\mathcal{C}_{g,m}|=k$.
For later use, we define the aggregate data-active and data-inactive subcarrier sets over one OFDM-IM block as
\begin{equation}
\mathcal{K}_{a} \triangleq \bigcup_{g=1}^{G} \mathcal{K}_{g,a}, \qquad \mathcal{K}_{p} \triangleq \bigcup_{g=1}^{G} \mathcal{K}_{g,p},
\end{equation}
with cardinalities
\begin{equation}
K_a = |\mathcal{K}_{a}| = Gk, \qquad K_p = |\mathcal{K}_{p}| = G(n-k).
\end{equation}
In conventional OFDM-IM, the subcarriers in $\mathcal{K}_{g,p}$ are not activated with data. In the proposed scheme, these data-inactive subcarriers are reused for artificial-noise (AN) transmission. Therefore, they are inactive only with respect to information-bearing data symbols, but active as security and sensing resources.

\subsection{Transmit Signal Model}

For a data-active subcarrier $\ell\in\mathcal{K}_{a}$, Alice transmits a beamformed data symbol as
\begin{equation}
\mathbf{x}_{\ell} = \mathbf{w}_{\ell}s_{\ell}, \qquad \ell\in\mathcal{K}_{a},
\label{eq:active_signal}
\end{equation}
where $\mathbf{x}_{\ell}\in\mathbb{C}^{N_t\times 1}$ is the transmitted vector, $\mathbf{w}_{\ell}\in\mathbb{C}^{N_t\times 1}$ is the data precoding vector, and $s_{\ell}$ is an $M$-ary constellation symbol satisfying $\mathbb{E}[|s_{\ell}|^2]=1$. The corresponding data covariance matrix is
\begin{equation}
\mathbf{W}_{\ell} = \mathbb{E} \left[ \mathbf{x}_{\ell} \mathbf{x}_{\ell}^{H} \right] = \mathbf{w}_{\ell} \mathbf{w}_{\ell}^{H}, \qquad \ell\in\mathcal{K}_{a}.
\label{eq:W_def}
\end{equation}
For a data-inactive subcarrier $\ell\in\mathcal{K}_{p}$, Alice transmits AN according to
\begin{equation}
\mathbf{x}_{\ell} = \mathbf{q}_{\ell}, \qquad \mathbf{q}_{\ell} \sim \mathcal{CN} \left( \mathbf{0}, \mathbf{V}_{\ell} \right), \qquad \ell\in\mathcal{K}_{p},
\label{eq:passive_signal}
\end{equation}
where $\mathbf{V}_{\ell}\succeq\mathbf{0}$ is the AN covariance matrix. The rank-one AN representation $\mathbf{V}_{\ell}=\mathbf{v}_{\ell}\mathbf{v}_{\ell}^{H}$ is recovered as a special case when a single AN stream $\mathbf{q}_{\ell}=\mathbf{v}_{\ell}z_{\ell}$ is transmitted with $z_{\ell}\sim\mathcal{CN}(0,1)$.
Let $\mathbf{h}_{B,\ell}\in\mathbb{C}^{N_t\times 1}$ and $\mathbf{h}_{E,\ell}\in\mathbb{C}^{N_t\times 1}$ denote the frequency-domain channel vectors from Alice to Bob and Eve, respectively, on subcarrier $\ell$. We define the corresponding channel covariance matrices as
\begin{equation}
\mathbf{H}_{B,\ell} = \mathbf{h}_{B,\ell} \mathbf{h}_{B,\ell}^{H}, \qquad \mathbf{H}_{E,\ell} = \mathbf{h}_{E,\ell} \mathbf{h}_{E,\ell}^{H}.
\label{eq:channel_covariances}
\end{equation}
The received signal at Bob on a data-active subcarrier is
\begin{equation}
y_{B,\ell} = \mathbf{h}_{B,\ell}^{H} \mathbf{w}_{\ell}s_{\ell} + n_{B,\ell}, \qquad \ell\in\mathcal{K}_{a},
\label{eq:bob_active}
\end{equation}
where $n_{B,\ell}\sim\mathcal{CN}(0,\sigma_B^2)$ is additive noise at Bob. On a data-inactive subcarrier carrying AN, Bob receives
\begin{equation}
y_{B,\ell} = \mathbf{h}_{B,\ell}^{H} \mathbf{q}_{\ell} + n_{B,\ell}, \qquad \ell\in\mathcal{K}_{p}.
\label{eq:bob_passive}
\end{equation}
To preserve Bob's OFDM-IM index detection capability, the AN covariance is designed such that the received AN leakage at Bob is bounded as
\begin{equation}
\operatorname{tr} \left( \mathbf{H}_{B,\ell} \mathbf{V}_{\ell} \right) \leq \epsilon, \qquad \ell\in\mathcal{K}_{p},
\label{eq:bob_null}
\end{equation}
where $\epsilon>0$ is a small leakage tolerance.
At Eve, the received signal on a data-active subcarrier is
\begin{equation}
y_{E,\ell} = \mathbf{h}_{E,\ell}^{H} \mathbf{w}_{\ell}s_{\ell} + n_{E,\ell}, \qquad \ell\in\mathcal{K}_{a},
\label{eq:eve_active}
\end{equation}
where $n_{E,\ell}\sim\mathcal{CN}(0,\sigma_E^2)$. On a data-inactive subcarrier, Eve receives
\begin{equation}
y_{E,\ell} = \mathbf{h}_{E,\ell}^{H} \mathbf{q}_{\ell} + n_{E,\ell}, \qquad \ell\in\mathcal{K}_{p}.
\label{eq:eve_passive}
\end{equation}
This design makes the received AN power on data-inactive subcarriers comparable to the received data power on data-active subcarriers at Eve. Consequently, an energy-based index detector at Eve becomes less able to distinguish true data-active subcarriers from AN-filled data-inactive subcarriers.

\subsection{Target-Direction Transmit Power}

Let $\mathbf{a}(\theta_t)\in\mathbb{C}^{N_t\times 1}$ denote the BS transmit steering vector toward the target direction $\theta_t$. The corresponding target steering matrix is defined as
\begin{equation}
\mathbf{A}_{t} = \mathbf{a}(\theta_t) \mathbf{a}^{H}(\theta_t).
\label{eq:A_t}
\end{equation}
The aggregate transmit covariance matrix over one OFDM-IM block is
\begin{equation}
\mathbf{R}_{X} = \sum_{\ell\in\mathcal{K}_{a}} \mathbf{W}_{\ell} + \sum_{\ell\in\mathcal{K}_{p}} \mathbf{V}_{\ell}.
\label{eq:total_covariance}
\end{equation}
Hence, the target-direction transmit power is
\begin{equation}
P_{\mathrm{T}}(\theta_t) = \operatorname{tr} \left( \mathbf{A}_{t} \mathbf{R}_{X} \right) = \sum_{\ell\in\mathcal{K}_{a}} \operatorname{tr} \left( \mathbf{A}_{t} \mathbf{W}_{\ell} \right) + \sum_{\ell\in\mathcal{K}_{p}} \operatorname{tr} \left( \mathbf{A}_{t} \mathbf{V}_{\ell} \right).
\label{eq:radar_power}
\end{equation}
Unlike ZF-based secrecy methods that reduce the signal power toward Eve, the proposed design maintains target-direction transmit power and instead protects the OFDM-IM index bits by making Eve's index observations ambiguous.

\subsection{Energy-Based Index Detection}

For an OFDM-IM receiver, the active-index pattern is detected independently in each subblock by selecting one legal codeword from $\mathcal{C}_{g}$. Given the received subcarrier observations, an energy-based index detector estimates the active pattern in the $g$-th subblock as
\begin{equation}
\hat{m}_{r,g} = \arg\max_{m\in\{1,\ldots,2^{p_1}\}} \sum_{\ell\in\mathcal{C}_{g,m}} \left| \tilde{y}_{r,\ell} \right|^2, \qquad r\in\{B,E\},
\label{eq:index_detector}
\end{equation}
where $\tilde{y}_{r,\ell}$ denotes the received observation at receiver $r$ on subcarrier $\ell$. If $m_g$ is the transmitted index-pattern label in subblock $g$, the corresponding index detection error event is defined as
\begin{equation}
\hat{m}_{r,g} \neq m_g.
\end{equation}
In the proposed system, Bob is expected to achieve reliable index detection because the AN leakage on data-inactive subcarriers is suppressed through \eqref{eq:bob_null}. In contrast, Eve's detector is intentionally degraded because the AN-filled data-inactive subcarriers are designed to resemble data-active subcarriers in received energy.

\section{Proposed AN-Aided OFDM-IM ISAC Design and Optimization Problem Formulation}
\label{sec:proposed_design}
In this section, we formulate the proposed AN-aided OFDM-IM ISAC design. The objective is to preserve legitimate OFDM-IM index detection at Bob, mislead Eve's index detector, and maintain strong target-direction transmit power.

\subsection{Index-Deception Constraint}

Let the average received data-active subcarrier power at Eve be defined as
\begin{equation}
\bar{P}_{E,a} = \frac{1}{K_a} \sum_{j\in\mathcal{K}_{a}} \operatorname{tr} \left( \mathbf{H}_{E,j} \mathbf{W}_{j} \right).
\label{eq:eve_avg_active_power_subblock}
\end{equation}
The proposed index-deception strategy forces the received AN power on each data-inactive subcarrier to be close to $\bar{P}_{E,a}$. Specifically, for $\ell\in\mathcal{K}_{p}$, we impose
\begin{equation}
\left| \operatorname{tr} \left( \mathbf{H}_{E,\ell} \mathbf{V}_{\ell} \right) - \bar{P}_{E,a} \right| \leq \delta, \qquad \ell\in\mathcal{K}_{p},
\label{eq:eve_matching_absolute}
\end{equation}
where $\delta>0$ is the allowed energy-matching tolerance. This constraint makes the AN-filled data-inactive subcarriers resemble data-active subcarriers to increase the ambiguity of Eve's energy-based index detector.

\subsection{Optimization Problem}
The proposed transmit covariance design maximizes the target-direction transmit power in \eqref{eq:radar_power} while satisfying Bob's receive SNR requirement, Bob-side AN suppression, Eve-side index-deception, and transmit power constraints. The optimization problem is formulated as
\begin{subequations}
\label{eq:main_opt_problem}
\begin{align}
\max_{\{\mathbf{W}_{\ell}\},\{\mathbf{V}_{\ell}\}} \quad & \sum_{\ell\in\mathcal{K}_{a}} \operatorname{tr} \left( \mathbf{A}_{t} \mathbf{W}_{\ell} \right) + \sum_{\ell\in\mathcal{K}_{p}} \operatorname{tr} \left( \mathbf{A}_{t} \mathbf{V}_{\ell} \right) \label{eq:opt_obj} \\
\text{s.t.} \quad & \operatorname{tr} \left( \mathbf{H}_{B,\ell} \mathbf{W}_{\ell} \right) \geq \Gamma\sigma_B^2, \quad \ell\in\mathcal{K}_{a}, \label{eq:opt_bob_qos} \\
& \operatorname{tr} \left( \mathbf{H}_{B,\ell} \mathbf{V}_{\ell} \right) \leq \epsilon, \quad \ell\in\mathcal{K}_{p}, \label{eq:opt_bob_an_null} \\
& \left| \operatorname{tr} \left( \mathbf{H}_{E,\ell} \mathbf{V}_{\ell} \right) - \bar{P}_{E,a} \right| \leq \delta,  \quad \ell\in\mathcal{K}_{p}, \label{eq:opt_eve_match} \\
& \sum_{\ell\in\mathcal{K}_{a}} \operatorname{tr} \left( \mathbf{W}_{\ell} \right) + \sum_{\ell\in\mathcal{K}_{p}} \operatorname{tr} \left( \mathbf{V}_{\ell} \right) \leq P_{\max}, \label{eq:opt_total_power} \\
& \operatorname{tr} \left( \mathbf{W}_{\ell} \right) \leq P_{\ell}^{\max}, \quad \ell\in\mathcal{K}_{a}, \label{eq:opt_sub_power_w} \\
& \operatorname{tr} \left( \mathbf{V}_{\ell} \right) \leq P_{\ell}^{\max}, \quad \ell\in\mathcal{K}_{p}, \label{eq:opt_sub_power_v} \\
& \mathbf{W}_{\ell} \succeq \mathbf{0}, \quad \ell\in\mathcal{K}_{a}, \label{eq:opt_psd_w} \\
& \mathbf{V}_{\ell} \succeq \mathbf{0}, \quad \ell\in\mathcal{K}_{p}. \label{eq:opt_psd_v}
\end{align}
\end{subequations}
Here, $\Gamma>0$ is the required receive SNR threshold at Bob for the data-active subcarriers, $\epsilon>0$ controls the allowable AN leakage at Bob, $P_{\max}>0$ is the total transmit power budget, and $P_{\ell}^{\max}>0$ is the per-subcarrier power limit.
Constraint \eqref{eq:opt_bob_qos} guarantees reliable data-symbol reception on the active subcarriers at Bob. Constraint \eqref{eq:opt_bob_an_null} suppresses the AN received by Bob on data-inactive subcarriers to preserve the OFDM-IM index structure. Constraint \eqref{eq:opt_eve_match} is the key index-deception constraint, which forces the AN observed by Eve on data-inactive subcarriers to resemble the average received data-active subcarrier energy. The total and per-subcarrier power constraints are imposed in \eqref{eq:opt_total_power}-\eqref{eq:opt_sub_power_v}, while \eqref{eq:opt_psd_w} and \eqref{eq:opt_psd_v} enforce positive semidefinite transmit covariance matrices.
The absolute-value constraint in \eqref{eq:opt_eve_match} is equivalent to two affine inequalities and therefore preserves convexity after the rank-one beamforming constraints are relaxed.

\subsection{Solution via Semidefinite Relaxation}

Problem~\eqref{eq:main_opt_problem} is solved in the covariance domain as an SDR of the original rank-constrained beamforming problem. The non-convex rank-one constraint on $\mathbf{W}_{\ell}$ is omitted, yielding a convex SDP over $\mathbf{W}_{\ell}\succeq 0$ and $\mathbf{V}_{\ell}\succeq 0$. If a rank-one data beamformer is required, it can be obtained from the principal eigenvector of $\mathbf{W}_{\ell}$ or by Gaussian randomization. The AN covariance $\mathbf{V}_{\ell}$ may remain higher-rank, since multidimensional AN can better satisfy the Bob-side suppression and Eve-side index-deception constraints.



\section{Numerical Results}
\label{sec:numerical_results}

In this section, we evaluate the proposed AN-aided OFDM-IM ISAC scheme and compare it with a classical ZF-secured OFDM baseline. Unless otherwise stated, the BS is equipped with $N_t=8$ antennas and the target/Eve is located at $\theta_t=20^{\circ}$. The OFDM-IM waveform uses $N=8$ subcarriers in the SDR-based optimization, where $K_a=4$ subcarriers are data-active and $K_p=4$ subcarriers are data-inactive but AN-bearing. The transmit power budget is swept from $10$~dBm to $40$~dBm. Bob's channel is modeled as a frequency-selective Rayleigh fading channel, while Eve/target is modeled as a line-of-sight channel. The required SNR threshold at Bob is set to $\Gamma=10$~dB, null space tolerance at Bob is $\epsilon=10^{-5}$, and allowed energy matching tolerance at Eve is $\delta=10^{-3}$. The SDR problems are solved using CVX, a MATLAB-based modeling system for convex optimization \cite{cvx}, and averaged over independent Bob-channel realizations.

\begin{figure}[!t]
    \centering
    \includegraphics[width=1\linewidth]{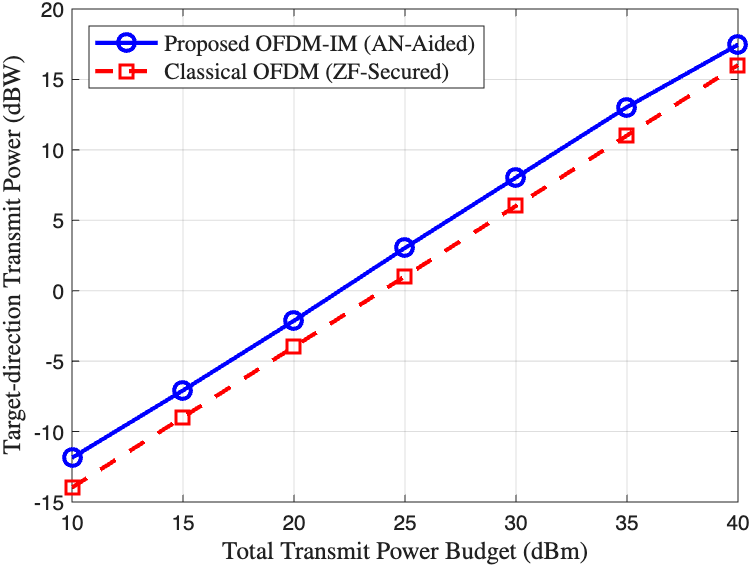} 
    \caption{Average transmit power towards the target versus the total transmit power budget.}
    \label{fig:radar_power}
\end{figure}

For comparison, we evaluate a classical ZF-secured OFDM baseline where all subcarriers are data-bearing. The ZF transmit covariance matrices $\{\mathbf{W}_{k}^{\mathrm{ZF}}\}$ are obtained by maximizing the transmit power toward the target $\sum_{k=1}^{N} \operatorname{tr}(\mathbf{A}_{t}\mathbf{W}_{k}^{\mathrm{ZF}})$ subject to Bob's SNR constraint $\operatorname{tr}(\mathbf{H}_{B,k}\mathbf{W}_{k}^{\mathrm{ZF}}) \geq \Gamma\sigma_B^2$ for all $k$, a total power limit $P_{\max}$, and Eve-directed spatial nulling $\operatorname{tr}(\mathbf{H}_{E,k}\mathbf{W}_{k}^{\mathrm{ZF}}) \leq \epsilon$ on a selected subset of subcarriers $\mathcal{K}_{\mathrm{ZF}}$ (where $|\mathcal{K}_{\mathrm{ZF}}| = K_p$). Fig.~\ref{fig:radar_power} compares the average transmit power toward Eve of the proposed AN-aided OFDM-IM scheme and the ZF-secured OFDM baseline as a function of the total transmit power budget. The proposed scheme achieves higher transmit power toward Eve over the considered power range. This result follows from the fact that the proposed method does not suppress the signal toward Eve. Instead, it transmits AN on the data-inactive subcarriers and shapes this AN to appear as active-subcarrier-like energy at Eve. Therefore, the AN contributes positively to transmit power toward Eve while the ZF-secured OFDM baseline reduces the power available for it. This demonstrates a fundamental ISAC security tradeoff: protecting communication by suppressing Eve can degrade sensing when the eavesdropper is also the target. The proposed method avoids this tradeoff by replacing power suppression with index deception.

\begin{figure}[!t]
    \centering
    \includegraphics[width=1\linewidth]{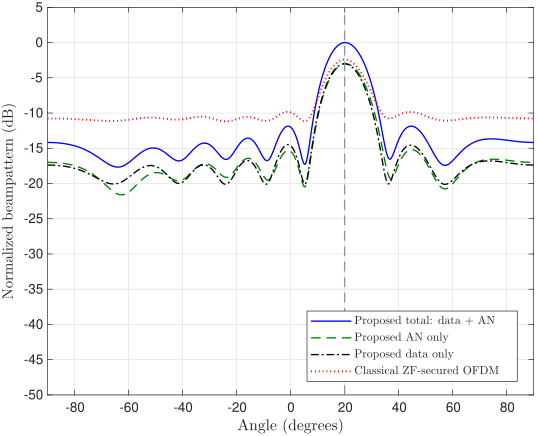} 
    \caption{Normalized spatial beampattern comparison with $\theta_t = 20^{\circ}$.}
    \label{fig:beampattern}
\end{figure}

Fig.~\ref{fig:beampattern} shows the normalized spatial beampatterns for a representative channel realization. The proposed total beampattern points toward the target/Eve direction, and the AN-only component also contributes significant target-direction power. In contrast, the ZF-secured OFDM baseline reduces power toward Eve to provide secrecy.


\begin{figure}[!t]
    \centering
    \includegraphics[width=1\linewidth]{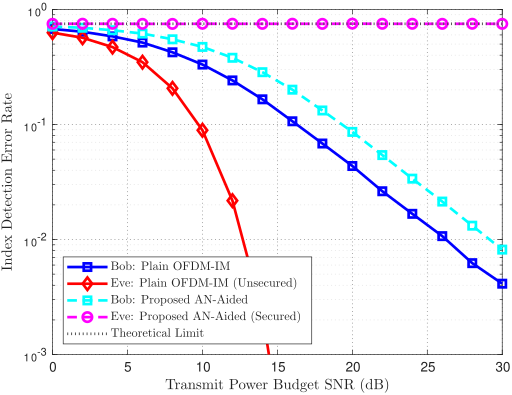} 
    \caption{Index detection error rate for Bob and Eve.}
    \label{fig:index_bler}
\end{figure}


Fig.~\ref{fig:index_bler} presents the index detection performance of Bob and Eve. In this simulation, $n=4$, $k=2$, and $2^{p_1}=4$ index patterns are used. As a result, the random-guessing index error probability for Eve is
$P_{\mathrm{Eve}} = 1-\left(\frac{1}{4}\right) = 0.75$. For plain OFDM-IM, Eve's index detection error rate decreases with SNR, showing that the active index pattern can be recovered through energy-based index detection when the inactive subcarriers are AN-free. This confirms the index-domain vulnerability of conventional OFDM-IM. In the proposed AN-aided scheme, however, Eve's index error rate remains close to the random-guessing limit because the AN-filled data-inactive subcarriers have active-like received energy. As a result, Eve cannot reliably distinguish true active subcarriers from deceptive AN-bearing subcarriers. Bob's index error rate decreases with SNR for both plain and proposed OFDM-IM. The proposed Bob curve exhibits an SNR shift compared to plain OFDM-IM due to the fact that part of the transmit power is allocated to AN instead of data.




\section{Conclusion}
\label{sec:conclusion}

This paper proposed an AN-aided secure OFDM-IM ISAC framework for scenarios where the sensing target may also act as an eavesdropper. An SDR-based transmit covariance design was formulated to maximize target-direction transmit power under Bob-side reliability, AN suppression, index-deception, and power constraints. Numerical results showed that data-inactive OFDM-IM subcarriers can be reused as dual-functional resources that preserve target-direction transmit power while degrading Eve's index detection.


\bibliographystyle{IEEEtran}

\bibliography{references}

\end{document}